\documentclass{JAC2000}

\usepackage{graphicx}

\begin{document}
\title{PRODUCTION AND STUDIES OF PHOTOCATHODES FOR HIGH INTENSITY ELECTRON BEAMS}

\author{E. Chevallay, S. Hutchins, P. Legros, G. Suberlucq, H. Trautner, CERN, Geneva, Switzerland}

\maketitle

\begin{abstract}
For short, high-intensity electron bunches, alkali-tellurides
have proved to be a reliable photo-cathode material.
Measurements of lifetimes in an RF gun of the CLIC Test
Facility II at field strengths greater than 100 MV/m are
presented. Before and after using them in this gun, the
spectral response of the Cs-Te and Rb-Te cathodes were
determined with the help of an optical parametric oscillator.
The behaviour of both materials can be described by Spicer's
3-step model. Whereas during the use the threshold for
photo-emission in Cs-Te was shifted to higher photon energies,
that of Rb-Te did not change. Our latest investigations on the
stoichiometric ratio of the components are shown. The
preparation of the photo-cathodes was monitored with 320 nm
wavelength light, with the aim of improving the measurement
sensitivity. The latest results on the protection of Cs-Te
cathode surfaces with CsBr against pollution are summarized.
New investigations on high mean current production are
presented.
\end{abstract}

\section{Introduction}
In the CTF II drive beam gun, Cs-Te photocathodes are used to
produce a pulse train of 48 electron bunches, each 10\,ps long
and with a charge of up to 10\,nC \cite{linac99}. In CTF, the
main limit to lifetime is the available laser power, which
requires a minimal quantum efficiency (QE) of 1.5\,\% to
produce the nominal charge. Although Cs-Te photocathodes are
widely used, a complete understanding, especially of their
aging process, is still lacking. Spectra of the QE against
exciting photons may help to understand the phenomenon.

\section{Measurements of QE against photon energy}
According to Spicer \cite{spicer1}, the spectra of the quantum
efficiency (QE) of semiconductors with respect to the energy of the
exciting photons ($h\nu$) can be described as:
\begin{equation}\label{eq:spicer}
  QE=\frac{c_1 (h\nu -E_T)^\frac{3}{2}}{c_2 + (h\nu
  -E_T)^\frac{3}{2}},
\end{equation}
where $E_T$ is the threshold energy for photoemission, c$_1$
and c$_2$ are constants.

\subsection{The OPO}
\begin{figure}[htb]
\centering\framebox{
\includegraphics*[width=60mm]{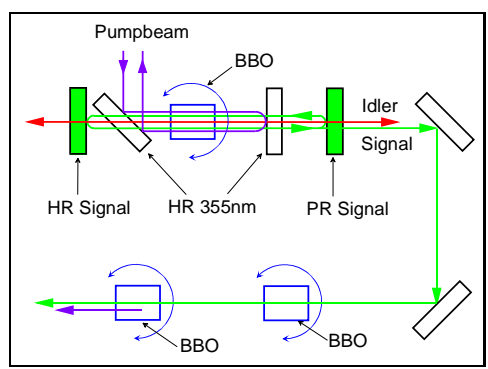}}
\caption{OPO scheme with following doubling crystals}
\label{fig:opo}
\end{figure}

To measure the spectral response of photocathodes, wavelengths
from the near UV throughout the visible are necessary. To
attain these, an {\bf O}ptical {\bf P}arametrical {\bf
O}scillator was built \cite{doc}. A frequency-tripled Nd:YAG
laser pumps a BetaBarium Borate (BBO) crystal in a double-pass
configuration, as shown in Fig.\,\ref{fig:opo}. The emerging
signal-beam, with wavelengths between 409\,nm and 710\,nm, is
frequency doubled in two BBO crystals. The wavelengths
obtained are between 210\,nm and 340\,nm. The idler-beam
delivers wavelengths between 710\,nm and $2600\,$nm.

The measurements of the spectral response of photocathodes
were made in the DC-gun of the photoemission lab at CERN
\cite{labor}, at a field strength of about 8 MV/m. Spectra
were taken shortly after the evaporation of the cathode
materials onto the copper cathode plug, as well as after use
in the CTF II RF-gun \cite{gun} at fields of typically 100
MV/m.

\subsection{Cesium Telluride}
\begin{figure}[htb]
\centering\framebox{
\includegraphics*[width=60mm]{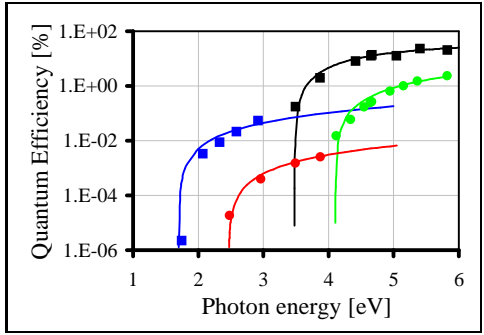}}
\caption{Spectra of a Cs-Te photocathode on copper. Before
(square points) and after use (round points) in the CTF II
drive beam gun} \label{fig:cath87}
\end{figure}

To be able to interpret the spectra in terms of Spicer's
theory, it was necessary to split the data into 2 groups, one
at ``low photon energy'' and one at ''high photon energy'',
see Fig.\,\ref{fig:cath87}. Then, the data can be fitted well
with two independent curves, following Eq.\,(\ref{eq:spicer}),
which give two threshold energies. For a typical fresh Cs-Te
cathode, the high energy threshold is 3.5\,eV, the low one is
1.7\,eV, as shown in Fig.\,\ref{fig:cath87}, upper curve.
This might be a hint that two photo-emissive phases of Cs-Te
on copper exist. Several explanations are possible: The copper
might migrate into the Cs-Te, creating energy levels in the
band gap; or possibly not only Cs$_2$Te, but also other Cs-Te
compounds might form on the surface and these might give rise
to photoemission at low photon energy. A hint to this might be
that the ratio of evaporated atoms of each element is not
corresponding to Cs$_2$Te, see below.

After use, we found that not only the complete spectrum
shifted towards lower quantum efficiency, but also that the
photoemission threshold for high QE increased to 4.1\,eV,
which is shown in Fig.\,\ref{fig:cath87}, lower curve. One
might expect that the photocathode is poisoned by the residual
gas, preventing low-energy electrons from escaping. However,
because typical storage lifetimes are of the order of months,
the effect must be connected to either the laser light, or the
electrical field.

\begin{figure}[htb]
\centering\framebox{
\includegraphics*[width=60mm]{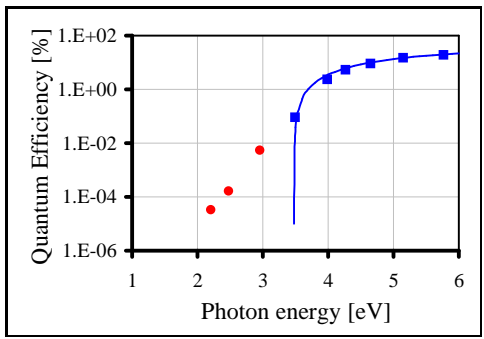}}
\caption{Cs-Te cathode, evaporated on a gold film of 100\,nm
thickness. The round points are not used for the fit.}
\label{fig:cath120}
\end{figure}

We also produced a Cs-Te cathode on a thin gold film of
100\,nm thickness. As shown in Fig.\,\ref{fig:cath120}, the
shoulder in the low energy response disappeared. It is
difficult to fit a curve for the Spicer model to the low
energy data. The ``high" photoemission threshold is at
3.5\,eV. At the moment, this cathode is in use in the CTF II
gun and will be remeasured in the future. In terms of
lifetime, this cathode is comparable to the best Cs-Te
cathodes, as it has already operated for 20 days in the
RF-gun.

\subsection{Rubidium Telluride}
\begin{figure}[htb]
\centering\framebox{
\includegraphics*[width=60mm]{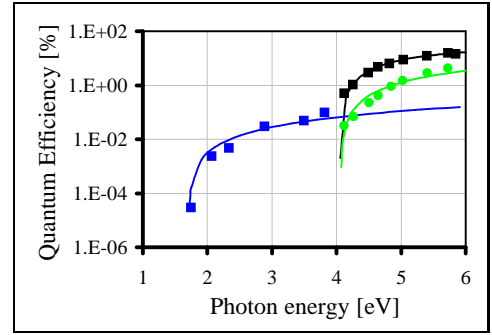}}
\caption{Spectra of a Rb-Te cathode before use (square points)
and after use (round points) in the CTF II drive beam gun}
\label{fig:rb2te}
\end{figure}

As a new material presented first in \cite{linac99}, we tested
rubidium-telluride. We took spectra of QE before and after use
in the CTF II gun, as for Cs-Te. Remarkably, with this
material, there was no shift in the photoemission threshold
towards higher energies, but only a global shift in QE, see
Fig.\,\ref{fig:rb2te}. This might be due to the lower affinity
of rubidium to the residual gas. Detailed investigations are
necessary to clarify this.

\subsection{Coating with CsBr}
\begin{figure}[htb]
\centering\framebox{
\includegraphics*[width=60mm]{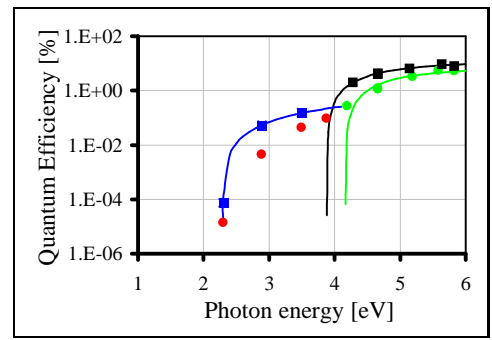}}
\caption{Spectra of a Cs-Te cathode without CsBr (square
points) and with CsBr coating (round points)} \label{fig:csbr}
\end{figure}

Long lifetimes for Cs-Te cathodes are achieved only when they
are held under UHV ($\leq 10^{-8}$\,mbar). Other photocathode
materials like K-Sb-Cs are immunized against gases like oxygen
by evaporating thin films of CsBr onto them \cite{csbr1}.
Therefore, we evaporated a CsBr film of 2\,nm thickness onto
the Cs-Te. Fig.\,\ref{fig:csbr} shows the spectrum before the
CsBr film (square points) and after it (round points).
The QE at 266\,nm dropped from 4.3\,\% to 1.2\,\%. In
addition, the photoemission threshold was shifted from 3.9\,eV
to 4.1\,eV. A long-term storage test showed no significant
difference between uncoated and coated cathodes. More
investigations will determine the usefulness of these
protective layers.

\subsection{Preparation with other wavelengths}
In order to increase the sensitivity of the on-line QE
measurement during evaporation of the photocathodes, we
monitored the process with light at a wavelength of 320\,nm.
We did not see any significant improvement in sensitivity,
notably in the high QE region.

\section{Stoichiometric Ratio} Film thicknesses are measured
during the evaporation process by a quartz oscillator
\cite{labor}. Typical thicknesses for high quantum
efficiencies at $\lambda=266\,$nm are 10\,nm of tellurium and
around 15\,nm of cesium. This results in a ratio of the number
of atoms of each species of $N_{Te}/N_{Cs}=2.85$, far from the
stoichiometric ratio of 0.5 for Cs$_2$Te. It is known that
tellurium interacts strongly with copper \cite{cste4}, so that
not all of the evaporated tellurium is available for a
compound with subsequently evaporated cesium. Therefore, we
used also Mo and Au as substrate material. However, the ratio
between the constituents necessary for optimum QE, did not
change significantly. Another reason might be that instead of
Cs$_2$Te,  Cs$_2$Te$_5$ is catalytically produced on the
surface. This compound, as well as some others, was found to
be stable \cite{cste1}.

\section{Lifetime in CTF II}
\begin{figure}[htb]
\centering\framebox{
\includegraphics*[width=60mm]{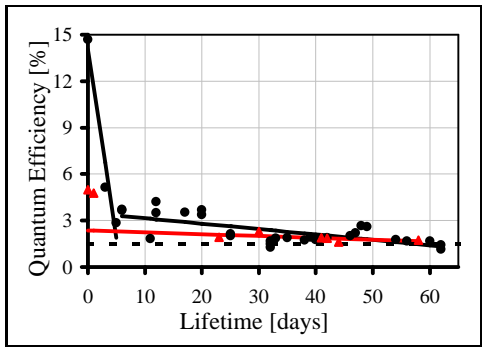}}
\caption{Lifetime in CTF of two different cathodes}
\label{fig:lifetime}
\end{figure}

Lifetime in CTF depends on parameters like maximum field
strength on the cathode, vacuum and especially extracted
charge. Typically, a cathode is removed from the gun, if the
QE falls below 1.5\,\%. As shown in Fig.\,\ref{fig:lifetime},
lifetime does not depend on the initial QE; a cathode having
an initial QE of 15\,\% (round points) lasted as long as one
with 5\,\% (triangles).

\section{High charge test}
As shown in Table\,\ref{tab1}, the average current produced in
CTF II is nearly a factor 10000 lower than what is required
for the CLIC drive beam. A test to produce 1\,mC is under
preparation in the photoemission laboratory at CERN. The exact
reproduction of the CLIC pulse structure would require the
CLIC laser, which is still in the design stage
\begin{table}[htb]
\begin{center}
\caption{Comparison of cathode relevant parameter}
\begin{tabular}{|c|c|c|c|}
\hline
          & CTF II    & CLIC 3\,TeV & "Test 1\,mC" \\ \hline
  Current & 0.008\,mA & 75\,mA      & 1\,mA     \\ \hline
  Power   & 0.072\,mW & 35\,W       & 300\,mW    \\ \hline
\end{tabular}
\label{tab1}
\end{center}
\end{table}
in a collaboration between Rutherford Appleton Laboratory and
CERN. A test which is compatible with our current installation
is the production of 1\,mA of dc current, which requires a UV
laser power of 300\,mW at the cathode. For this test, we will
illuminate the cathode with pulses of 100\,ns to 150\,ns pulse
length, at repetition rates between 1\,kHz and 6\,kHz. As
Table\,\ref{tab1} shows, this is a factor 1000 more average
current than in CTF II, and also demonstrates the basic
ability of the cathodes to produce the CTF 3 drive beam
(I=26$\mu$A). CLIC is still a factor 75 away. We are currently
searching for ways to produce higher charges as well.

\section{Conclusion}
Measurements of QE against photon energy are routinely made
after production and after use of photocathodes. We have
demonstrated that both low energy and high energy responses
agree well with Spicer's theory. A gold buffer layer reduces
the low energy response of Cs-Te cathodes. More work is needed
to understand the measurements of the stoichiometric ratio of
Cs-Te. Coating with 2\,nm CsBr significantly decreased the
quantum efficiency, without improving the storage lifetime.
For the high-charge drive beam of CLIC, it is still necessary
to demonstrate the capabilities of Cs-Te, for which first
tests will be done soon.

\end{document}